# Investigating student understanding of bipolar junction transistor circuits

Kevin L. Van De Bogart and MacKenzie R. Stetzer

*Department of Physics and Astronomy, University of Maine, Orono, ME 04469*

**Abstract**

The research reported in this article represents a systematic, multi-year investigation of student understanding of the behavior of bipolar junction transistor circuits using a variety of different tasks to isolate and probe key aspects of transistor circuit behavior. The participants in this study were undergraduates enrolled in upper-division physics electronics courses at three institutions, as well as undergraduates in upper-division engineering electronics courses at one of the institutions. Findings from this research indicate that many students have not developed a robust conceptual understanding of the functionality of bipolar junction transistors circuits even after all relevant instruction. Most notably, when asked to analyze the impact of a transistor circuit on input signals, students frequently applied reasoning appropriate for an analysis of the circuit's dc bias behavior. However, students often displayed knowledge of fundamental transistor behavior when responding to more targeted questions. This article provides insight into student thinking about transistor circuits, describing the most prevalent conceptual and reasoning difficulties identified and discussing some important implications for instruction.

## I. INTRODUCTION

A large and substantive body of research on the learning and teaching of physics at the undergraduate level has been conducted in the context of introductory courses (see for example, [1–8]). In the last decade or so, a growing number of researchers have taken interest in student understanding of physics content beyond the introductory level; leading to work in upper-division courses on mechanics [9], electricity and magnetism [10], quantum mechanics [11], and thermodynamics [12]. Relatively recently, researchers have begun focusing on upper-division laboratory instruction [13–17].

Upper-division electronics courses are a common element of undergraduate physics programs, and typically focus on either analog electronics or a hybrid of analog and digital electronics. These courses represent an intersection between content not typically taught elsewhere in the undergraduate physics curriculum and skill-based learning goals associated with laboratory instruction (including, for example, the development of troubleshooting expertise). Indeed, the *AAPT Recommendations for the Undergraduate Physics Laboratory Curriculum* identify the development of laboratory skills as one of six critical focus areas for laboratory courses [18]. Several recent and ongoing efforts have investigated both the content and skills developed through electronics instruction, including investigations of student understanding of operational-amplifier circuits [19] and student troubleshooting [20–22].

Despite this increased attention on upper-division electronics, student understanding of the content covered in these courses has not yet been studied in sufficient depth to inform large-scale, research-based instructional improvements in treatment of key electronics circuits and concepts. Such efforts are particularly important given that there is reason to believe that students do not enter these courses with a robust understanding of basic circuits. Indeed, research by McDermott and Shaffer revealed that students often fail to develop a coherent conceptual model for simple DC circuits after traditional instruction in introductory calculus-based physics courses [23]. The authors identified several conceptual and reasoning difficulties, and used their findings to develop research-based and research-validated instructional materials on electric circuits [24]. Subsequent research by Engelhardt and Beichner associated with the development of the DIRECT (a research-based instrument focused on introductory circuits content) also yielded similar findings [25]. A recent investigation probing student understanding of circuits in both introductory and advanced courses revealed that students in upper-division electronics courses struggled with foundational circuits concepts, including Kirchhoff's junction rule and the notion of a complete circuit [26].

While electronics courses are ubiquitous in contemporary physics and engineering curricula, the research base on the learning and teaching of content covered in such courses is rather limited. Over the past decade, some work has been conducted by physics education researchers and engineering education researchers on student understanding of phase behavior in ac circuits [27], filters [28,29], and operational-amplifier circuits [19,30].

To date, however, there has been no published work specifically focused on student conceptual understanding of the behavior of bipolar junction transistor (BJT) circuits. While there has been some work outline strategies for teaching transistor circuits in engineering education journals, it has primarily centered around a pedagogical approach of combining transistors into functional groups [31,32]. These articles do not provide any data on the efficacy of such an approach, nor do they provide insight into the kinds of difficulties students might encounter when analyzing basic circuits involving a single transistor.

In order to better understand which aspects of transistor behavior are well understood by students after instruction and which are sources of ongoing difficulties, we conducted an in-depth investigation across multiple institutions. By providing a comprehensive description of student understanding of transistor circuits, this work establishes a solid research base that may inform instruction on the topic as well as the development of targeted, research-based instructional materials.

The study described in this paper was guided by the following research questions:

1. To what extent do students develop an understanding of basic bipolar junction transistor circuits after relevant instruction in an electronics course?
2. To what extent do students demonstrate that they understand the functional role of transistors in typical circuit applications?

3. What ideas and approaches, both correct and incorrect, do students employ when analyzing transistor circuits?

In order to address these questions, we developed and administered several research tasks to students enrolled in four upper-division electronics courses in physics and engineering at three different institutions. In this paper, we limit our discussion to a total of five research tasks, which focus on different aspects of transistor circuit behavior. A sixth related task targeting the behavior of ac biasing networks (which are frequently included in transistor circuits) has been detailed in a separate publication [33].

We begin with a brief overview of the research context and methodology (Sec. II). In Secs. III-V, we present the individual research tasks along with the associated results and insights into student thinking. In Sec. VI, we discuss student difficulties identified as well as implications for instruction, and we summarize our findings in Sec. VII.

## II. RESEARCH CONTEXTS AND METHODS

This investigation of student understanding of bipolar junction transistor circuits was conducted in courses at three different universities. In this section, we provide a brief overview of the courses studied, the relevant instruction on bipolar junction transistors, and the methods associated with the investigation.

### A. Upper-division electronics courses studied

Data for this investigation were collected in laboratory-based electronics courses at three different four-year public research universities (denoted U1, U2, and U3). The specific courses in which the investigation was conducted are characterized in Table I. Research tasks were administered in physics electronics courses at all three institutions (with each course covering a similar spectrum of topics), as well in as one engineering course at U1.

All courses were required for their respective majors, were typically taken during the first semester of the junior year, required the submission of formal lab reports, and included written exams on circuit analysis. While the experiments in the physics electronics courses at U1 and U2 were usually completed within the allocated laboratory time, the experiments associated with the engineering electronics course at U1 and the physics electronics course at U3 typically required additional work in the laboratory outside of the official course hours. Finally, the course at U3 was unique in that students were required to give a presentation on a 5-week group project at the end of the semester.

**Table I.** Overview of courses in which data were collected.

| Institution | U1 | | U2 | U3 |
|---|---|---|---|---|
| Course Discipline | Physics | Engineering | Physics | Physics |
| Textbook | Diefenderfer, Galvez, or Lawless | Sedra and Smith | Horowitz and Hill | Horowitz and Hill |
| Enrollment (Students) | 10-20 | 25-45 | 30-80 | 30-60 |
| Laboratory Time (Hr/wk) | 2 | 3 | 3 | 3 |
| Lecture Time (Hr/wk) | 2 | 3 + 1.5 Recitation | 2 | 2 |

**B.  Brief overview of BJT circuit coverage**

In all four courses, students are taught that a bipolar junction transistor (BJT) may act as a current amplifier under appropriate conditions, such as those provided by the circuit depicted in

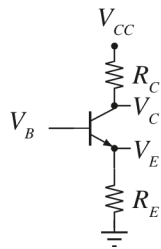

**Fig. 1.** The canonical BJT common-emitter amplifier circuit. The base, collector, and emitter voltages have been labeled ($V_B$, $V_C$, and $V_E$ respectively) for clarity.

Fig. 1. The three terminals of a BJT are the collector, base, and emitter, with voltages (with respect to ground) denoted $V_C$, $V_B$, and $V_E$, respectively, in Fig. 1. In a first-order model of the BJT as a current amplifier, the relationship between these three voltages determines the operational mode of the transistor, which in turn determines the relationships between currents at each junction. Below, we paraphrase the requisite conditions for an *npn* transistor to be forward-active, as articulated by Horowitz and Hill [34]

1. The voltage at the collector ($V_C$) must be higher than the voltage at the base ($V_B$), which must in turn be higher than the voltage at the emitter ($V_E$).

2. The base-emitter junction behaves similarly to a pn diode, and will conduct current when $V_{BE} = V_B - V_E$ is approximately +0.6 V.

When both of these conditions are satisfied, $I_C$ is roughly proportional to $I_B$ and can be written as $I_C = \beta I_B$, where $\beta$ is typically about 100. In the forward-active regime, the base and collector currents, $I_B$ and $I_C$, are directed into the transistor, and the emitter current, $I_E$, is directed out of the device. From Kirchhoff's junction rule, $I_E = (\beta+1) I_B$, which is frequently approximated as $I_E \approx I_C$.

While this is a relatively informal treatment of transistors, when combined with Kirchhoff's Laws, it is sufficient for making predictions about the behavior of BJT circuits operating in the forward-active regime. Although some courses covered more sophisticated models of BJT behavior, students in all courses were expected to be able to use the simple model described above to explain the basic behavior of emitter follower circuits (*i.e.,* circuits with an output voltage measured at the emitter, or $V_E$ in Fig. 1) as well as common-emitter amplifier circuits (*i.e.,* circuits with an output voltage measured at the collector, or $V_C$ in Fig. 1)

C. **Methodology**

Since the primary goal of this work was to document student thinking in sufficient detail to inform efforts to improve instruction on these topics, this investigation was designed and conducted through the lens of the specific difficulties empirical framework [35–37]. While there was a considerable emphasis on the specific conceptual and reasoning difficulties students encounter, we also sought to identify those aspects of transistor circuit analysis at which students were generally successful.

In order to elicit ideas about transistor circuits from students, we administered written free-response research tasks that explicitly prompted students to explain their reasoning. These tasks were given as ungraded conceptual questions or were included on course exams. As a result, students typically had a limited amount of time to respond to these questions (generally 10-15 minutes).

While the answers given by students were usually unambiguous and supported by a small assortment of explanations, the exact wording of each explanation was unique to each student. Thus, it was necessary to perform further analysis of students' reasoning in order to generalize responses sufficiently for broader characterization. To this end, a grounded theory approach [38,39] was employed to identify the general lines of reasoning used from the specific responses provided by students.

**III. THREE AMPLIFIER COMPARISON TASK**

The common-emitter amplifier is used extensively for small-signal voltage amplification, and represents an important building block employed in more complex circuits. Given the ubiquity of the common-emitter amplifier circuit in both instruction and electronics applications, our first research task was designed to probe student understanding of this relatively complex but important circuit. As in previous investigations [19], we sought to elicit student thinking by asking students to analyze and compare three amplifier circuits, two of which represent slight modifications from a base circuit (shown in Fig. 2 and drawn from Ref. [34]).

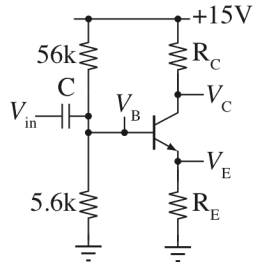

**Fig. 2.** Canonical BJT common-emitter amplifier circuit used in the three amplifier comparison task. The base ($V_B$), collector ($V_C$), and emitter ($V_E$) voltages have been labeled for clarity.

### A. Task Overview

In the three amplifier comparison task, shown in Fig. 3, students must compare the small-signal behavior of three transistor circuits, all of which are properly biased. Students are explicitly told the left portions of all three circuits were identical. Circuit B is a standard common-emitter amplifier circuit, and the other two are slight modifications of circuit B. In circuit A, the collector and emitter resistors are switched, which affects the amplifier's gain. In circuit C, the output terminal is at the emitter, and thus the circuit is an emitter follower. Students are asked to rank, from largest to smallest, the peak-to-peak amplitudes of the output voltages of all three circuits ($V_{out,A}$, $V_{out,B}$, and $V_{out,C}$) and to explain their reasoning.

In order to answer this question correctly, students need a sufficiently robust understanding of BJT circuit behavior in order to ascertain the impact of switching the collector and emitter resistors (B *vs.* A) as well as the impact of switching the location of the output terminal (B *vs.* C). It should be noted that instructors felt that the task was reasonable to ask of students in their course; some explicitly indicated that students should be able to answer the task correctly after BJT instruction in their courses.

B. **Correct Response**

In this section, we provide the correct reasoning in terms of gain expressions, followed by an overview of the reasoning required to derive the three circuits' behavior from first principles. In all courses, students would have seen or derived ac gain equations for the common-emitter amplifier and follower circuits; students were not expected to rederive the gain equations for this task. When discussing the circuit behavior in detail, there is a need to differentiate between signals (or variations in voltage) and dc voltages; a prefix of a lower-case delta (δ) is used when discussing periodic variations in voltages with respect to time, whereas dc quantities are presented with no prefix.

Since circuits A and B are both properly biased common-emitter amplifiers, their peak-to-peak amplitudes are therefore given by the gain expression $\delta V_{out} = -\delta V_{in}(R_C/R_E)$. This result implies that for the given component values, $|\delta V_{out,A}| = 1/10\, \delta V_{in}$ and $|\delta V_{out,B}| = 10\, \delta V_{in}$. For circuit C, also biased appropriate, the emitter follower, $\delta V_{out,C} = \delta V_{in}$ (*i.e.*, the gain is 1), so the correct ranking of all three peak-to-peak voltages is $|\delta V_{out,B}| > |\delta V_{out,C}| > |\delta V_{out,A}|$.

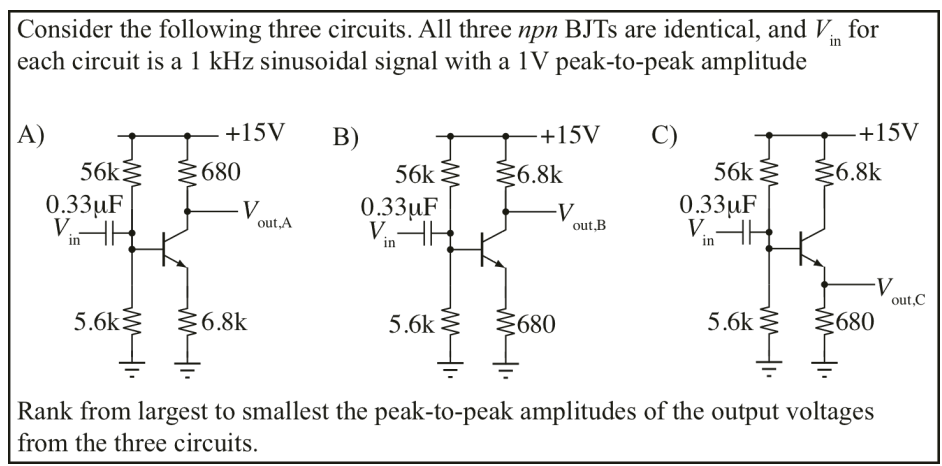

**Fig. 3.** Three amplifier comparison task. Note that the text has been slightly abridged/paraphrased and that explanations were required.

*More detailed analysis of common-emitter amplifier task.* Transistor circuits typically require input signals to be biased around a constant, non-zero dc voltage in order to function properly. The left three components in all three circuits (the "blocking" capacitor $C$, the 56-k$\Omega$ resistor, and the 5.6-k$\Omega$ resistor) form a biasing network that serves both to remove any existing dc offset from $V_{in}$ (the primary function of the capacitor) as well as to introduce a new, constant offset of +1.36 V. The biasing network is also equivalent to a biased high-pass filter designed such that it does not attenuate amplitude of the input signal provided (*i.e.*, $\delta V_B = \delta V_{in}$ for the 1-kHz signal in this case). Student analysis of ac biasing networks is discussed in more detail in Ref. [33].

Since the ac biasing networks are identical in all three circuits, the given 1 V peak-to-peak amplitude of the input voltage ($\delta V_{in} = 1$ V) will result in a base voltage in all three circuits that is also 1 V peak-to-peak ($\delta V_B = 1$ V) and centered about a +1.36 V dc offset.

Since $\delta V_B = 1$ V and the signal has a +1.36 V dc offset, the transistor is always properly biased (*i.e.*, $V_{BE} = 0.6$ V). As a result, $V_E = V_B - 0.6$ V and the peak-to-peak amplitudes at the base and emitter are necessarily the same, or $\delta V_E = \delta V_B = \delta V_{in}$. This final expression characterizes the essential small-signal behavior of emitter follower circuits (*e.g.*, circuit C in Fig. 3).

Knowing that $I_E = V_E/R_E$ (from Ohm's law) and $I_E \approx I_C$, the collector's voltage behavior can be determined as $V_C \approx 15$ V $-(V_E/R_E) R_C$ and therefore the variation is collector voltage is $\delta V_C = - (R_C/R_E) \delta V_E$. Thus, since $\delta V_E = \delta V_B = \delta V_{in}$, $\delta V_C = - (R_C/R_E) \delta V_{in}$, which characterizes the essential small-signal behavior of common-emitter amplifier circuits.

## C. Overview of student performance

We collected data from this task in the engineering electronics course at U1 (N = 57) as well as in the physics electronics courses at U1 (N = 42), U2 (N = 169), and U3 (N = 142), with a total of 410 student responses. The question was administered either as an ungraded conceptual question or as part of a final exam. To ensure that results may be presented in a concise fashion, the more formal notation of variations and absolute value has been omitted.

As shown in Table II, the distribution of students giving a correct ranking of the peak-to-peak output voltages for all three circuits ($V_{out,B} > V_{out,C} > V_{out,A}$) varied widely across different courses, ranging from 7% to 42% of the students in a given course. Furthermore, less than a third of students in any given course gave correct reasoning in support of a completely correct answer, and nearly all of these students used the common-emitter amplifier gain expressions (without deriving them). Thus, it is evident that the majority of students struggled on this task.

The most common incorrect ranking, given by approximately one third of students in each course, was that $V_{out,A} > V_{out,B} > V_{out,C}$. One student supported this ranking in the following manner:

> "Because in circuit A, there won't be as much of a voltage drop across the 680 Ω resistor as there will be across the 6.8 kΩ resistor. As circuits B and C have the voltage divider switched, A will be greater. And $V_B > V_C$ due to the voltage drop across the transistor."

Here, the student made the comparison between circuits A and B by first considering the collector and emitter resistors and then ranking the output of circuit A as larger than that of B due to the smaller resistor causing a smaller voltage drop with respect to the +15V supply. Such a response implicitly (and incorrectly) assumes that the collector currents in the two circuits are the same. The comparison between B and C was made with the idea that there was a decrease in voltage from the collector to the emitter of the transistor, and thus the output of circuit B ($V_{out,B}$) must be at a higher voltage than the output of circuit C ($V_{out,C}$). Both of these approaches are critical for analyzing the bias (dc) voltages in the circuit, but neither is sufficient for comparing peak-to-peak voltages.

Between 10 and 40% of all students in a given course appeared to be ranking dc voltages by providing the kind of qualitative reasoning described above. In addition, between 5 and 30% of students in each course used formal calculations of the bias voltages in order to rank the output voltages of the three circuits. Combined, these two methods of comparing dc voltages account for between 16% and 30% of the reasoning provided by students in a given course.

The next most prevalent ranking was $V_{out,B} > V_{out,A} > V_{out,C}$, given by between 10% and 30% of students. In support of this ranking, one student wrote:

> "... [T]he voltage drop across $V_{out,B}$ will be the largest because there is a larger resistance between it and the +15 V. So, ratings go $V_{out,C} < V_{out,A} < V_{out,B}$."

While such responses differed greatly in the justification given for ranking circuit C (with no prevalent commonalities), they all used the same sort of reasoning to compare circuits A and B;

**Table II.** Overview of student performance on the transistor amplifier comparison task (shown in Fig. 3).

| Context | U1 Eng | U1 Phy | U2 Phy | U3 Phy |
|---|---|---|---|---|
| Number of responses | 57 | 42 | 169 | 142 |
| $V_B > V_C > V_A$ **(Correct)** | 7% | 14% | 14% | 42% |
| *correct & complete reasoning* | *0%* | *2%* | *6%* | *33%* |
| $V_A > V_B > V_C$ | 33% | 38% | 30% | 32% |
| *dc reasoning* | *16%* | *33%* | *20%* | *29%* |
| $V_B > V_A > V_C$ | 30% | 10% | 12% | 9% |
| $V_B > V_A$ | 46% | 31% | 38% | 58% |
| *correct & complete reasoning* | *4%* | *9%* | *11%* | *49%* |
| $V_{out} \propto R_C$ | *7%* | *7%* | *10%* | *2%* |
| $V_B < V_A$ | 51% | 60% | 53% | 39% |
| *dc reasoning* | *21%* | *33%* | *32%* | *12%* |
| $V_B > V_C$ | 77% | 73% | 65% | 86% |
| *correct & complete reasoning* | *0%* | *3%* | *6%* | *33%* |
| *dc reasoning* | *7%* | *21%* | *22%* | *10%* |

namely, they argued that the larger collector resistor in circuit B results in a larger voltage (*i.e.*, they compared voltages across resistors). While a correct ranking of the peak-to-peak amplitudes relies on the ratio of the resistances of the collector and emitter resistors, the written responses suggested that these students were implicitly assuming that the currents in both circuits are identical.

**D. Pairwise Comparisons**

While the overall ranking and reasoning used provide valuable insight into students' thinking about transistor circuits, it was evident that many students struggled when comparing the outputs of all three circuits simultaneously. Thus, it was useful to examine how students treated the relevant modification to the canonical base circuit: either changing the location of the output terminal (circuit B *vs.* circuit C) or exchanging the collector and emitter resistors (circuit B *vs.* circuit A). In practice, few students specifically compared circuits A and C, and furthermore few students stated that any of the circuits would have outputs equal to one another. A summary of the breakdown of student comparisons of the output voltages from each pair of circuits is presented in Table II.

There was substantial variation in students' comparisons of circuits B and A, with the correct ranking ($V_{out,B} > V_{out,A}$) given by between 30% and 60% of students in a given course. A total of 5% to 50% of students in a given course correctly supported this ranking using either the circuit gains of the amplifiers or the ratio of collector to emitter resistors. A smaller number of students (between 2 and 10% of a given course) justified the same comparison using arguments based solely on the collector resistors rather than the ratio. This may be due to students confusing the voltage across the collector resistor with the circuit's output voltage.

A sizable number of students responded that $V_{out,B} < V_{out,A}$, accounting for between 40% and 60% of responses from a given course. For this ranking, the most prevalent single line of reasoning was to state that the voltage drop from +15 V would be proportional to the collector resistance, which accounted for between 12% and 33% of responses from a given course. Such responses did not attend to any variations in the collector current over time (and also never addressed the fact that collector currents for a given input voltage differ between the two circuits) and are thus implicitly focused on dc behavior.

Of students who compared circuits B and C, approximately 75% correctly found that $V_{out,C} < V_{out,B}$, with a spread between 66% and 86% of students giving correct comparisons in each course. However, this particular comparison can stem from both a ranking of peak-to-peak (signal) amplitudes and a ranking of (dc) bias voltages. This is reflected in the reasoning used by students, with only between 0% and 33% of students in a given course comparing the small-signal gains of the circuits. Furthermore, no more than one third of students in any course explicitly identified C as a follower circuit.

The next most common line of reasoning, provided by between 7% and 22% of students from a given course, was to instead conclude that the output of B is greater than that of C by reasoning that there would be a voltage drop across the transistor, which supports a comparison of dc parameters. While this response is consistent with the application of salient knowledge about transistor behavior, it is not productive for an analysis of the peak-to-peak amplitudes of the circuit output voltages.

### E. Specific Difficulties Identified

The three prevalent circuit rankings discussed above account for at least half of all responses from any given institution. This suggests that the difficulties associated with this task are likely to be relevant to most electronics courses. Several specific difficulties were identified.

***Failure to differentiate between signal and dc bias.*** After all instruction on transistors, the majority of students gave responses that did not address the small-signal behavior of the circuits. Instead, the most common lines of reasoning were appropriate only for finding dc voltages. This suggests that many students failed to differentiate between the small-signal and dc behavior of the circuits; moreover, the relatively poor performance across all institutions suggests that students did not recognize that the characteristic behaviors of the common-emitter amplifier and the emitter follower apply to signals – not bias voltages.

***Tendency to use single, local features to make comparisons.*** The most common incorrect lines of reasoning employed comparisons based primarily on the relative resistances of the resistors adjacent to the output terminals, as was observed in between 10% to 40% of student responses from each course. Such approaches may be a manifestation of the kind of local reasoning strategies exhibited in introductory circuits [23], and they also support the hypothesis that even upper-division students may not yet have developed coherent models of circuits.

***Failure to recognize the intended functionalities of the common-emitter amplifier and the emitter follower.*** The relatively poor performance on this task across all institutions suggests that students did not recognize that the characteristic behaviors of the common-emitter amplifier and the emitter follower apply to the *signals* – not the bias voltages. Indeed, for both kinds of circuits, students tended to apply analysis strategies suitable for dc voltages. In particular, students' reasoning pertaining to the follower circuit varied wildly, but typically did not focus on its intended functionality. This observation is consistent with the fact that less than one third of students in any course clearly identified circuit C as a follower.

**F.     Discussion**

In all four courses, the majority of students were unable to correctly rank the peak-to-peak output voltages of the follower and common-emitter amplifier circuits in this task, even after all instruction on transistors. Indeed, roughly half of students in each course struggled to correctly compare the peak-to-peak output voltages from just the two amplifier circuits A and B. This is surprising, as all instructors judged this task to be reasonable for students in their course. Students frequently considered only the dc behavior of the circuit, which was not aligned with the functionality of either the common-emitter amplifier or the emitter follower. It is important to note that all of the identified difficulties were exhibited by students at all institutions; although students at U3 were generally more successful on the task, over half were unable to rank the three output voltages correctly.

## IV. APPLICATIONS OF BASIC TRANSISTOR MODEL

On the three amplifier comparison task, the majority of the students struggled in their efforts to rank all three circuits according to the amplitudes of the peak-to-peak output voltages. Given the complexity of this task, it is hard to use student responses to pinpoint specific difficulties associated with the functionality of the transistor itself. Indeed, students frequently adopted reasoning that did not clearly draw upon properties of the transistor. Moreover, even the most common line of correct reasoning (comparison via gain equations) did not require discussion of the transistor's electrical properties. Thus, it is difficult to tell from the three amplifier comparison task if students possessed an unarticulated understanding of basic transistor properties or if they were unaware of how transistors functioned in general. As a result, several additional tasks were developed that examined how well students understood the fundamental behaviors of BJTs.

### A. Follower Current Ranking Task

The first of these tasks was designed to probe the extent to which students were able to recognize the fundamental relationships among the three terminal currents when the transistor is in the forward-active regime.

#### 1. Task Overview

In the follower current ranking task (see Fig. 4), students are shown a simple BJT emitter follower circuit, in which the input voltage (at the base of the transistor) is +3 V, the collector is connected to a +15 V supply, and the emitter is connected to ground via a 3.3-kΩ resistor. Although the prompt did not explicitly state that all components are ideal, in practice the students treated them as such. Students were asked to rank the currents through the base, collector, and emitter terminals of the transistor (labeled *X*, *W*, and *Y*, respectively), to state explicitly if any currents were equal or were equal to zero, and to explain their reasoning.

#### 2. Correct Response

To answer this question correctly, students must first recognize that for the selected voltages, the transistor will be in the forward-active regime. As a result, the transistor current gain equation may be applied ($I_C \approx 100\, I_B$) to compare the collector and base currents ($I_C > I_B$). Furthermore, as a consequence of Kirchhoff's current law, $I_E = I_C + I_B$, and thus the emitter current is necessarily the largest ($I_E > I_C$). In practice, as discussed in Section II.B, it is often assumed that $I_E \approx I_C$. Thus, in terms of the variables used in the prompt, a correct response would be either $I_Y > I_W > I_X$ or $I_Y = I_W > I_X$.

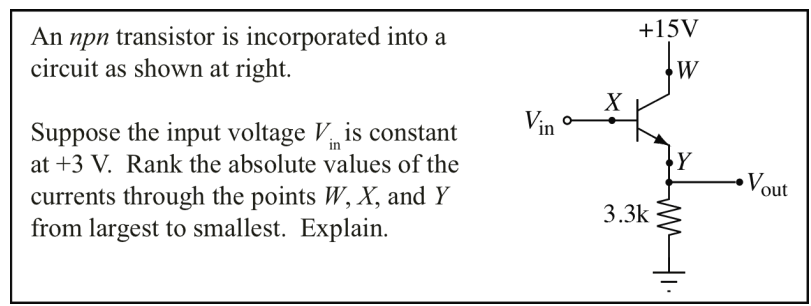

**Fig. 4.** Follower current ranking task.

### 3. Student Performance

The task was administered to a smaller cohort of students than the amplifier comparison task, corresponding to a single physics class at $U_1$ (N = 12) and three physics classes at $U_2$ (N = 155). Overall, students were considerably more successful on the follower current ranking task than they were on the three amplifier comparison task. However, only between 50% and 75% of students in a given class indicated the correct ranking of currents, and at least two-thirds of such students also provided correct reasoning in support of their ranking. Notably, for this task, no individual incorrect ranking accounted for more than 10% of the total number of responses given by students.

From this task, it is apparent that many students demonstrate some understanding of the functional relationships among the currents in BJT circuits under forward-active conditions. Given that previous research conducted in upper-division electronics courses revealed student difficulties with the application of Kirchhoff's junction rule in some situations [19,26], student performance on the follower current ranking task suggests that a significant percentage of students are, in fact, able to either apply the junction rule to simple BJT circuits successfully or draw upon a model of BJT behavior consistent with the junction rule.

### B. Follower Voltage Graphing Task

While the follower current ranking task probed the extent to which students understand the functional relationships among currents in a forward-active transistor, we also sought to ascertain the extent to which students could productively apply ideas about the voltage across the base-emitter (BE) junction in transistor follower circuits. To first order, the BE junction of an *npn* transistor may be treated as a diode in that there is no current through the junction before a 0.6 V threshold voltage is reached, and thereafter the current is determined by the circuit configuration. To better understand how students treat the BE junction when analyzing BJT circuits, the follower voltage graphing task (shown in Fig. 5) was created and administered.

### 1. Task Overview

In the follower voltage graphing task, students are presented with the same BJT follower circuit used in the follower current ranking task. For the new task, students are told that the input voltage increases linearly from -2 V to 2 V over a time interval of 8 seconds. Students are asked to produce a quantitatively correct graph of the output voltage in the space provided and to explain their reasoning.

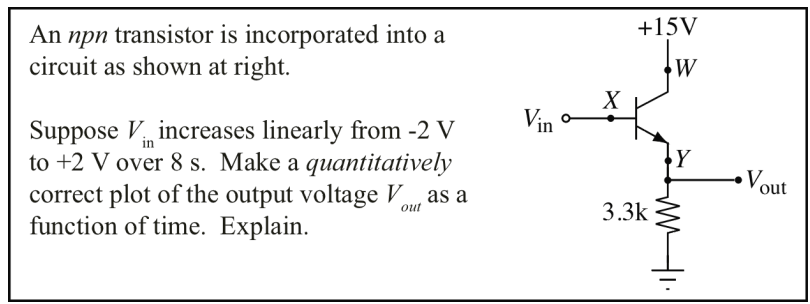

**Fig. 5.** Follower voltage graphing task. Note that students were provided with both a plot of $V_{in}$ vs. time and a labeled grid in which to plot $V_{out}$ vs. time.

## 2. Correct Response

To give a correct response to this task, a student must identify the time interval in which the voltage at the base is at least 0.6 V greater than ground, and the interval in which it is not. When $V_B > 0.6$ V, the transistor will be in the forward-active regime, and the voltage at the emitter will be (approximately) 0.6 V lower than the base (*i.e.*, $V_E = V_B - 0.6$ V). Otherwise, the diode-like base-emitter junction will not be forward biased, implying that $V_{out}$ will be 0 V since there will be no current through the emitter resistor. Therefore, a quantitatively correct graph of $V_{out}$ remains at 0 V until $V_{in}$ is 0.6 V (at $t \approx 5.2$ s), and then increases linearly, with the same slope as $V_{in}$, after that time (as depicted in Fig. 6A).

## 3. Student Performance

The task was administered to students in courses at U1 (N=57) and U2 (N=157), either paired with the follower current ranking task or as an entirely independent question. As shown in Table III, between 25% and 60% of students in a given course were able to produce a graph with the requisite quantitative features (*i.e.*, $V_{out} = V_{in} - 0.6$ V when $V_{in} > 0.6$ V, and $V_{out} = 0$ V otherwise). An additional 10% of students in either course produced graphs that had

**Table III.** Overview of student performance on the follower voltage graphing task (shown in Fig. 5).

|  | U1 (N = 57) | U2 (N = 157) |
|---|---|---|
| **Quantitatively correct** | **61%** | **25%** |
| *correct reasoning* | *58%* | *19%* |
| Qualitatively correct | 12% | 10% |
| *correct reasoning* | *10%* | *8%* |
| Linear with offset | 12% | 22% |
| *transistor acts as a diode* | *12%* | *18%* |
| Linear without offset | 2% | 12% |
| *configuration is a follower* | *2%* | *6%* |

*qualitatively* correct features although they were not quantitatively correct. Such qualitatively correct graphs depicted outputs that were 0 V before some threshold input voltage and were linear afterwards, but these graphs either had an incorrect threshold voltage or an incorrect slope.

Almost all students (>85% in any course) supported a quantitatively or qualitatively correct graph with correct reasoning. For example, one student wrote, *"The voltage $V_{out}$ is equal to $I_y R$, $I_y$ varies w/ $V_{in}$ and further, $V_y - .6\ V = V_{out}$ but only when $V_{in}$ is above .6 V, thus the one follows from the other, staggered by .6 V."* While the language used was informal, this student correctly described the diode-like behavior of the transistor's base-emitter junction.

The most common incorrect response, given by approximately 20% of students, was to depict a linearly increasing output voltage offset from the input voltage by a constant, negative value (typically –0.6 V), as shown in Fig. 6B. These students typically focused on the diode-like voltage drop of the transistor exclusively; for example, one student wrote, *"$V_{out}$ is equal to the emitter voltage. The emitter voltage is 0.6 volts less than the base voltage… which is $V_{in}$. $V_{out} = V_{in} - 0.6\ volts$"*. Nearly all (>80%) of the students who drew such graphs provided similar justifications for their responses. As noted previously, the relevant transistor property for this task is the diode-like behavior of the BE junction, which would never allow (significant) current from the emitter to the base due to the orientation of the *pn* junction. However, the negative output voltages associated with these incorrect student responses would require (significant) currents from the emitter to the base (*i.e.,* current traveling in the wrong direction through the *pn* junction). Thus, these responses likely stem from overgeneralizing the forward-biased voltage behavior of diodes to the transistor's BE junction, without properly attending to biasing constraints.

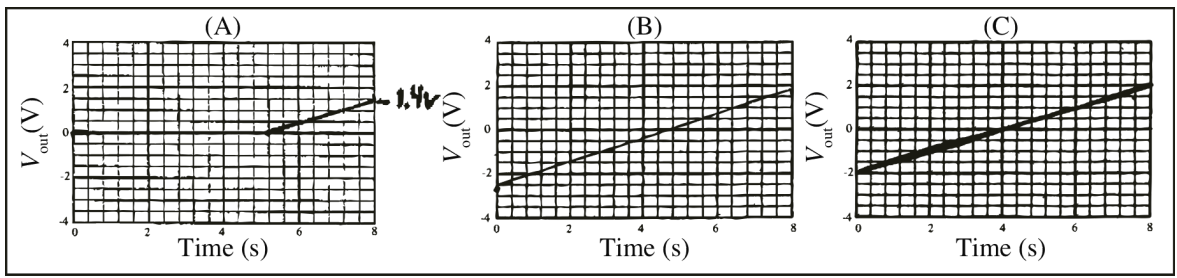

**Fig. 6.** Common student responses to the follower graphing task (shown in Fig. 5). These responses shown include: (A) a quantitatively correct graph, (B) a linear graph with an offset of –0.6 V, and (C) a linear graph with no offset.

The next most common incorrect response, given by roughly 10% of students, was to create a graph of the output that was identical to the input at all times, which is depicted in Fig. 6C. Approximately half of such responses were supported with reasoning similar to that written by the following student: *"Since $V_{in}$ increases linearly, $V_{out}$ has to increase linearly as well because there is nothing changing in the circuit."* Students may have correctly recognized the function of the circuit (*i.e.,* that the output voltage "follows" the input voltage) in such a response, but they appeared to assume that the output would follow exactly and unconditionally, which is not the case for an emitter follower.

The remaining responses given by students varied greatly, with no other answers accounting for more than 6% of student responses. Thus, the three categories of answers presented (as well as the associated lines of reasoning) fully characterize approximately two-thirds of responses given by students in either course. It should be noted that, even in common incorrect responses, most students were applying productive ideas about the behavior of transistors in their answers.

### 4. Specific Difficulties Identified

In this task, there were two common incorrect responses given by students, each of which had a strongly associated line of reasoning.

***Tendency to ascribe to the BE junction a fixed 0.6 V drop from $V_{in}$ to $V_{out}$ for all inputs.*** Between 12% and 22% of students in a given course treated the BE junction as having a fixed 0.6 V drop for the entire range of input voltages. Such responses did not attend to the biasing requirements for the transistor's forward-active behavior. Thus, even if students were considering the transistor to act as a diode, these responses did not capture the fact that semiconductor diodes must be forward biased by 0.6 V in order to allow current through the junction (from the anode to the cathode).

***Tendency to treat the BJT emitter follower as a dc follower with $V_{out} = V_{in}$ for all inputs.***
Approximately 10% of responses from U2 indicated that the output would be equal to the input, regardless of the value of $V_{in}$. These students did not address the biasing of the transistor in any way. It is likely that these students were effectively overgeneralizing follower behavior, failing to recognize that the emitter follower circuit is a signal follower ($\delta V_{out} = \delta V_{in}$) but not a dc follower; indeed, in operational amplifier follower circuits, $V_{out} = V_{in}$ for both ac and dc voltages.

### 5. Summary of Findings

While most students identified some key aspects of the circuit's behavior, less than two-thirds of students in any course were able to produce a quantitatively correct graph with correct reasoning. Slightly over one-third of students in any given course produced graphs that were quantitatively incorrect but that had elements of a correct response, including relevant aspects of diode-like behavior such as maintaining the slope of the input voltage or recognizing that there should be no output voltage for negative input voltages. While students often struggled to recognize the biasing constraints of BJT follower circuits, most displayed some evidence of understanding the general behavior of the follower circuit under basic dc conditions.

### C. Transistor Supply Voltage Modification Task

In the three amplifier comparison task, students frequently (and implicitly) assumed that collector currents were equal when comparing amplifier circuits. In order to better gauge student understanding of the functional relationship between supply voltages and the resulting currents in transistor circuits, a new task was created, as shown in Fig. 7. By focusing on a single aspect of transistor behavior, this task was designed to probe in greater detail student understanding of the causal relationships between currents and voltages in transistor circuits.

1. **Task Overview**

In the transistor supply voltage modification task, students are presented with one base circuit that is modified slightly in each of the two parts of the task, as shown in Fig. 7. For each part of the task, students are asked to determine how, if at all, the specified change in supply voltage will impact the collector current (through point $W$), and to explain their reasoning. In the first part of the task, students are told that the collector supply voltage $V_{CC}$ is decreased from +15 V to +10 V. In the second part of the task, students are told that the emitter supply voltage $V_{EE}$ is increased from 0 V to +1 V. It is important to note that such modifications were not an explicit part of instruction and are not typically discussed in detail in most texts; as a result, students

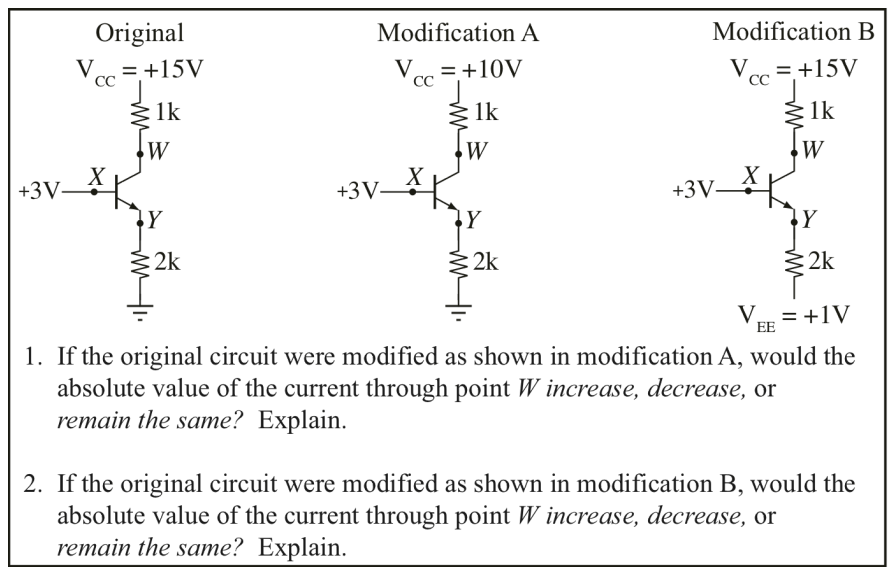

**Fig. 7.** Transistor supply voltage modification task. Note that the text has been abridged/paraphrased.

would be unlikely to give a memorized response.

## 2. Correct Response

In order to arrive at a correct response to either part of the task, students must recognize that the transistor remains forward-active in all cases, and thus the collector current is determined by the emitter current (since $I_C \approx I_E$), which is in turn set by the voltage drop across the emitter resistor $R_E$. In part 1 of the task, since $V_{in}$, $V_{EE}$, and the emitter resistor are unchanged, then both the emitter current (through point $Y$) and the collector current (through point $W$) remain the same. In part 2 of the task, increasing $V_{EE}$ results in a smaller potential difference across the 2-k$\Omega$ emitter resistor, which in turn decreases the collector current.

## 3. Student Performance

The task was administered to students in the physics electronics course at U1 (N = 27) over two different years. As shown in Table IV, the majority of students (78%) correctly recognized that changing the collector voltage $V_{CC}$ would not alter the current through point $W$ (*i.e.,* the collector current) in part 1 of the task, and nearly all of these students (74% of total) supported their answer with correct reasoning. For example, one student noted, *"If $V_{CC}$ were decreased to +10 V, the absolute value of the current through point W would stay the same since it is independent of $V_{CC}$. $I_C \approx I_E$."*

The remaining 22% of students all responded that the current would decrease, with their reasoning typically stating that the reduced voltage would translate into less current through the resistors in the circuit. For instance, one student responded as follows: *"W would decrease because there is less voltage to drop across the resistors. I = V/R, so with R constant and V decreased, I must decrease. (15 > 10)."* While the student is correct in reasoning that the current through a resistor should change if the voltage across that resistor changes (due to Ohm's law), this student did not recognize that in this case, the potential difference between the collector and emitter terminals *(i.e., between W and Y)* would vary in such a way that the emitter and collector currents remain essentially constant.

For the second part of the task, nearly all (~90%) students recognized that increasing the emitter voltage would subsequently decrease the collector current. In addition, 80% of these students supported their answers with correct reasoning. For example, one student wrote, *"$V_Y$ would be the same, but voltage drop needed across 2k resistor would be smaller, so $I_Y$ would be smaller. Since $I_Y = I_W$, current through $I_W$ would decrease."* Thus, most students correctly recognized that the current through the emitter resistor would decrease, and furthermore that the collector current would also necessarily decrease.

## 4. Summary of Findings

Overall, nearly two-thirds (63%) of students gave fully correct answers with correct reasoning on both parts of the task. Thus, this task demonstrates that many students do, in fact, have an understanding of the causal relationships that determine emitter and collector currents, and can use them productively in appropriate conditions. However, it should be noted that the most prevalent line of incorrect reasoning (accounting for approximately one fifth of student responses) stemmed from students reasoning that changing the collector voltage would necessarily impact the collector current, likely arguing that a change in one part of the circuit should have an impact in that same part of the circuit (*i.e.*, they are using local reasoning [23]). If students were consistently using local reasoning on both parts of the task, then it would be expected that they would respond by saying that the current would decrease in the first part and remain the same in the second. In practice, only a single student did so. Indeed, students' stronger performance on the second part of the task suggests that they were better able to draw upon the *non-local* relationship between transistor currents in the second scenario in order to recognize that a change in one part of a transistor circuit may in fact impact a transistor current in a different part of that same circuit.

**Table IV.** Overview of student performance on the transistor supply voltage modification task (shown in Fig. 7).

|  | U1 (N = 27) |
|---|---|
| Part 1: Reduced collector supply voltage $V_{CC}$ | |
| **Same current (correct)** | **78%** |
| *correct reasoning* | *74%* |
| Decreased current | 22% |
| *Ohm's law for collector resistor* | *19%* |
| Part 2: Increased emitter supply voltage $V_{EE}$ | |
| **Decreased current (correct)** | **89%** |
| *correct reasoning* | *70%* |

## V. REVISED AMPLIFIER COMPARISON TASK

It was noted in Section III that students typically struggled with analyzing the ac signal behavior of the emitter follower and common-emitter amplifier circuits, and many students seemed to give responses consistent with the behavior of those circuits under dc conditions. Most students who did not invoke the relevant gain expression for the common-emitter amplifier were unable to reproduce the reasoning required to compare the peak-to-peak amplitudes of the output voltages in the three circuits. However, as seen in Section IV in the discussion of student performance on the follower current ranking task, follower voltage graphing task, and the supply voltage variation task, many students demonstrated a general understanding of the functional behavior of forward-active transistors. Collectively, such results suggest that the original amplifier comparison task may have been overwhelming for students, and that the complexity may have inhibited students from applying their understanding productively.

In order to better probe student understanding of common-emitter amplifiers, a new task with less overhead was designed, shown in Fig. 8. The new task was designed such that (a) students are only asked to consider the impact of one modification at a time, (b) students must explicitly consider both the small-signal and dc behavior of the same circuits, and (c) all of the circuits compared are common-emitter amplifiers. These modifications were made in an effort to better understand to what extent students were struggling to choose appropriate techniques to evaluate signal and bias voltages in transistor circuits and to streamline data interpretation.

A. Task Overview

In the revised amplifier comparison task, students are presented with three common-emitter amplifier circuits, labeled A, B, and C. Circuit B differs from A solely in that it has a larger

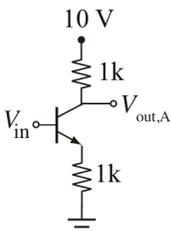

Three circuits with *npn* transistors are shown. $V_{in}$ is the same for all circuits.

$V_{in} = +2.6$ V.

1. Is $V_{out,B}$ *greater than*, *less than*, or *equal to* $V_{out,A}$? Explain.
2. Is $V_{out,C}$ *greater than*, *less than*, or *equal to* $V_{out,A}$? Explain.

$V_{in}$ is now a sinusoidal signal with a 1V pk-pk amplitude and a dc offset of +2.6 V.

3. Is the peak-to-peak amplitude of $V_{out,B}$ *greater than*, *less than*, or *equal to* that of $V_{out,A}$? Explain.
4. Is the peak-to-peak amplitude of $V_{out,C}$ *greater than*, *less than*, or *equal to* that of $V_{out,A}$? Explain.

**Fig. 8.** Revised amplifier comparison task. Note that the text has been abridged/paraphrased.

emitter resistor (2 kΩ *vs.* 1 kΩ), and circuit C differs from A solely in that it has a larger collector resistor (2 kΩ *vs.* 1 kΩ). Students are asked first to consider the dc behavior of the circuits, and to make pairwise comparisons between the output voltages from circuits B and A as well as those from C and A. For the next two parts of the task, students are asked instead to consider an appropriately biased signal, and to compare the peak-to-peak amplitudes of the output voltages of circuits B and A as well as those from C and A.

**B.   Correct Response**

In order to answer the (dc) portion of the task (parts 1 and 2) correctly, students must recognize that the emitter resistor determines the collector current for all three circuits. Since the emitter resistor in circuit B is larger than that in circuit A, the collector current in circuit B is less than that in A, resulting in a smaller voltage drop across the 1-kΩ collector resistor in circuit B, and thus there is a higher voltage at the output of circuit B than at the output of circuit A ($V_{out,B} > V_{out,A}$). In the second portion of the dc task, both circuits A and C have identical emitter resistors, but circuit C has a larger collector resistor than circuit A. Since the currents through the collector resistors in both circuits are also the same, there is a larger voltage drop across the collector resistor in circuit C. Thus, the output voltage of circuit C will be lower than that of circuit A ($V_{out,C} < V_{out,A}$).

For the corresponding signal portion of the task (parts 3 and 4), the magnitude of each circuit's small-signal gain (as noted in Section III) is given by the ratio of the collector resistor to the emitter resistor (*i.e.*, $\delta V_{out}/\delta V_{in} = R_C/R_E$). As such circuits A, B, and C have gains of 1, 2, and 1/2, respectively. Since all circuits have the same input signal, the peak-to-peak output voltage of circuit A is greater than that of circuit B ($\delta V_{out,A} > \delta V_{out,B}$) and the output voltage of circuit A is smaller than that of circuit C ($\delta V_{out,A} < \delta V_{out,C}$).

## C. Student Performance

Data were collected from the U1 physics electronics course over two different semesters (N = 29). Students were considerably more successful on this task than on the amplifier comparison task, as 50% of students answered all four parts correctly, and 86% of these students in turn also provided correct and complete reasoning for all parts, as shown in Table V. The remaining students who arrived at correct answers typically provided reasoning that contained correct elements, but was incomplete in some manner.

In responses to the four parts of the revised amplifier comparison task, 86% of students correctly answered both dc questions, and 64% correctly answered both signal questions. Furthermore, the majority of students supported their correct answers with correct reasoning. While it may initially appear that a larger percentage of students were able to correctly analyze the circuit for dc input voltages than for sinusoidal signal input voltages, a careful analysis of correct responses with correct reasoning indicates that there is not a statistically significant difference in performance on the two portions of the task ($p = .59$). While there is no evidence on this particular task to suggest that students struggle more to analyze circuits for signals than for constant dc input voltages, it is important to remember that the rather small sample size limits the claims that may be made.

On the original three amplifier comparison task in Section III, students struggled to compare the peak-to-peak output voltages of the two common-emitter amplifier circuits (A and B, in which the collector and emitter resistors were exchanged), with only 31% of U1 students across all years correctly ranking $V_{out,B} > V_{out,A}$, and only 30% of those students supporting their answers with correct reasoning. However, as was noted previously, 64% of students correctly described the small-signal behavior of both circuits in the revised task. While there are several important differences between the two tasks, the same reasoning is required in both cases. Using a $\chi^2$ test, there is a statistically significant difference in performance between the tasks ($p < .0001$, $\chi^2 = 20.45$) with a large effect size ($\Phi = .57$). This supports the hypothesis that students may in fact possess a relevant understanding of the small-signal behavior of common-emitter amplifier circuits, but they do not draw upon this relevant knowledge when answering the original amplifier comparison task; instead, they appear to draw upon dc approaches in the absence of an explicit juxtapositioning of the two analysis approaches. It is important to note that the request for a comparison of peak-to-peak amplitudes alone does not appear to be a sufficient prompt in the original three amplifier comparison task.

Given that students tended to use dc reasoning when prompted for peak-to-peak voltages in the original three amplifier comparison task, we examined the extent to which students used different approaches when asked to compare the same two circuits under dc and small-signal conditions (*e.g.,* part 1 *vs.* part 3 and part 2 *vs.* part 4). We found that 64% of students arrived at different answers for dc and small-signal conditions when comparing both pairs of circuits (B *vs.* A and C *vs.* A), and the majority of these students provided correct responses. This supports the idea that a significant percentage of students, when explicitly asked to consider both analyses,

**Table V.** Overview of student performance on the revised amplifier comparison task (shown in Fig. 8).

|  | U1 (N=29) |
|---|---|
| **All parts correct** | **50%** |
| *correct reasoning* | *43%* |
| Both dc questions correct | 86% |
| *correct reasoning* | *66%* |
| Both signal questions correct | 64% |
| *correct reasoning* | *55%* |

recognized the difference between dc and small-signal behavior.

However, 14% of students indicated that the comparisons between both pairs of circuits (B *vs.* A and C *vs.* A) were the same in both dc and small-signal cases. Moreover, these students employed the same reasoning across both kinds of comparisons. As a specific example, in response to the dc comparison in part 1, one student wrote: *"Since $I_E \approx I_C$, and there is less current in $V_{out,B}$ due to an increased resistance, I assume $V_{out,B} < V_{out,A}$."* In response to the analogous small-signal comparison in part 3, the same student wrote: *"$V_{out,B} < V_{out,A}$. The big factor, I believe again is the resistor in the E branch of circuit 2."* Such findings suggest that even after relevant instruction and explicit requests for two separate analyses, some students did not distinguish between the dc and small-signal behavior of the circuit and applied the same line of reasoning to both.

Care must be taken in the interpretation of these results, however, as they were obtained from a relatively small number of students. Nevertheless, these data are valuable in that they assist us in pinpointing what students can and cannot do and therefore help us to better interpret the poor student performance on the original three amplifier comparison task.

## VI. DISCUSSION

This investigation of student understanding of bipolar junction transistor circuits began with a single task (the three amplifier comparison task), which demonstrated that many students struggled to reason correctly in the context of a relatively common application of BJTs. In particular, students tended to use reasoning appropriate for dc quantities even when asked about small-signal properties. However, from responses to this task, it was unclear how well students understood the basic functionality and behavior of the transistors in such circuits, as the most common incorrect lines of reasoning did not explicitly address the behavior of the transistor itself. Through three additional, more targeted tasks (follower current ranking, follower voltage graphing, and transistor supply voltage modification), we found that students often did, in fact, demonstrate a basic understanding of transistor behavior. In the revised amplifier comparison task, which was considerably more scaffolded than the original three amplifier comparison task, students were significantly more successful in comparing the output signals of BJT circuits when students were explicitly asked to analyze the same circuits under both dc and small-signal conditions.

### A. Specific Difficulties Spanning Tasks

From the responses to these five tasks, students encountered a number of distinct difficulties when working with various BJT circuits. However, two overall trends emerged that were particularly noteworthy.

***Tendency to overgeneralize circuit element behavior.*** Across all tasks, students tended to make comparisons based on an overgeneralization of some property of the circuit elements. Typically, this consisted of using resistors as a proxy for voltages (or voltage drops) or always treating the voltage drop across a transistor junction as finite and constant. The former is only applicable for situations with identical currents, and the latter is only applicable for a forward-biased junction. Thus, the most common difficulties seem to result from a failure to apply appropriate constraints on circuit element behavior. Similar behavior has been observed for operational amplifier circuits, where students overgeneralized properties to conclude that there would be no current through any of the terminals of the op-amp [19].

***Tendency to rely on dc analysis over small-signal analysis.*** In instances in which students were not explicitly prompted to consider both dc and small-signal analyses of a circuit, students frequently used inappropriate strategies to reason about transistor circuits. As an example from the three amplifier comparison task, most incorrect lines of reasoning centered on arguments made about dc voltages, even though students were asked about peak-to-peak values of voltage signals. In addition, on the revised amplifier comparison task, even with explicit prompts for both analyses, about 15% of students still applied dc approaches when asked to compare the output voltage signals from two pairs of BJT circuits. Still, on the same task, roughly half of the students appeared to be capable of correctly predicting the small-signal behavior of the transistor when asked about both analyses explicitly and when presented with somewhat more straightforward circuits (*e.g.*, no biasing networks). Taken together, these results suggest that students may favor dc analysis over small-signal analysis, possibly because either they do not recognize that the small-signal behavior is relevant or they are less familiar with the appropriate analysis procedure for signals.

**B.     Implications for Instruction**

The findings from this research indicate that students do not develop a robust understanding of bipolar junction transistor circuits in typical electronics courses. On the basis of student performance on multiple research tasks, the combination of lecture instruction and laboratory experience provided in these courses does not appear to be sufficient for students to gain a thorough understanding of BJT functionality in many common circuits. However, there is evidence that the basic aspects of BJT behavior are relatively well understood. In addition, the most common incorrect lines of reasoning given by students still drew upon productive ideas about transistors. Our results suggest that there is a need for new instructional approaches to the content and for research-based instructional materials that can both leverage productive student ideas and address the difficulties identified in this investigation.

***Emphasize the relationship between the BE junction of the BJT and the semiconductor diode, including the constraints for current through each.*** Through the suite of research tasks described in this work, it has been shown that, in some contexts, many students could make accurate and well-reasoned predictions about the behavior of a transistor circuit. In particular, students were relatively adept at reasoning about the base-emitter junction's diode-like properties. Nonetheless, many students struggled to leverage the relationship between the BE junction and the semiconductor diode to recognize the biasing requirements for current and non-zero emitter voltages and to recognize that negative emitter voltages would require current to go "backwards" through the BE junction. An increased emphasis on the analogies between the two *pn* junctions and on the constraints associated with the behavior of each would likely improve student performance. Additional discussion of these constraints when covering diodes early in the course might also be beneficial.

***Guide students through both dc and small-signal analyses of common BJT circuits.*** Students often did not discriminate between the dc and small-signal behavior of common-emitter amplifier circuits in the absence of scaffolding supporting both analyses. Furthermore, they often favored dc analysis over small-signal analysis, even when the latter was required. As instructors, we often tend to focus on the intended functionality of specific circuits (*e.g.,* small-signal amplification for common-emitter amplifiers), and fail to help students recognize that the same circuit may behave very differently under different conditions (*e.g.,* dc input voltages). By asking student to analyze the behavior of given circuits for both dc and small-signal voltage inputs, students may be more likely to recognize the differences in both analysis approach and behavior. Such efforts should also help students differentiate between signal and bias.

***Help students juxtapose the behaviors of circuits with similar functionality comprised of different circuit elements.*** The first inverting and non-inverting op-amp amplifier circuits that many students encounter (and thus the first circuits with greater than unity gain) act identically on ac and dc voltages. Thus, it is possible that students who study op-amps before transistors (which is the case for some courses in this investigation) may generalize this behavior to transistor amplifiers as well. Giving students opportunities (in lab or in lecture) to contrast the behavior of op-amp amplifiers and transistor amplifiers (or op-amp followers and transistor followers) under different conditions may prevent students from overgeneralizing amplifier (or follower) behavior. It may also be productive to introduce circuits with asymmetric effects for dc voltages and sinusoidal signals (*e.g.*, op-amp amplifier circuits with dc biases) more frequently in the curriculum. Such an approach might also help students consider the role of biasing in these observed asymmetries.

## VII. CONCLUSION

This paper describes an in-depth investigation of student understanding of transistor circuits. On the three amplifier comparison task, the majority of students in each course were unable to correctly rank the three different transistor circuits according to peak-to-peak output voltage for identical input signals, and in most courses students struggled to support their responses with correct reasoning. The poor performance on this task was somewhat unexpected, as explicit instruction on these circuits (the common-emitter amplifier and the emitter follower) was included in all courses studied. Upon examining student responses, it was found that many students were not attempting to use (or derive) an appropriate small-signal gain expression for the circuits. Instead, students tended to reason about dc (bias) voltages in the circuit or used only a single relevant property of a circuit element in forming their response.

A series of additional tasks were created to better probe student understanding of more fundamental aspects of transistor behavior, as it was generally not possible to extract such information from the three amplifier comparison task. Students typically performed better on these more focused tasks than on the amplifier comparison task. Even after instruction, however, over one quarter of students were unable to correctly rank the terminal currents through a forward-active bipolar junction transistor on the follower current ranking task. Nevertheless, the additional tasks demonstrated that many students could indeed reason productively about transistor circuits from basic principles. Finally, results from the revised amplifier comparison task still suggest that approximately 15% of students fail to differentiate between dc analysis and small-signal analysis when making comparisons between various common-emitter amplifier circuits.

We are currently developing research-based instructional materials on BJT circuits in order to address many of the difficulties reported in this article. Findings from this investigation and others increasingly suggest that there may be a need for instructional approaches that span different classes of circuits. Namely, there appears to be a need to foreground: (1) circuits with similar functionality that are made from different components (*e.g.*, BJT followers vs. op-amp followers), and (2) similarities in behavior across circuit elements (*e.g.,* between the BE junction in BJTs and semiconductor diodes).

**ACKNOWLEDGMENTS**

We appreciate the substantive contributions made by colleagues in the UMaine Physics Education Research Laboratory, particularly for their input in framing and interpreting some of these results. We would also like to acknowledge feedback and input at various points during this project from Nuri Emanetoglu, Christos P. Papanikolaou, George S. Tombras, Lillian C. McDermott, Peter S. Shaffer, and Jason Alferness. In addition, we would like to thank the many colleagues who graciously allowed us to collect project data: Nuri Emanetoglu, H. J. Lewandowski, Miguel Morales, Donald Mountcastle, and Leslie J. Rosenberg. This material is based upon work supported by the National Science Foundation under Grant Nos. DUE-0618185, DRL-0962805, DUE-1022449, and DUE-1323426.